# Modern Symmetric Cryptography methodologies and its applications

Amin Daneshmand Malayeri
Department of Computer Engineering
Young Researchers Club, Malayer Azad University
Malayer, Iran
amin.daneshmand@gmail.com

Jalal Abdollahi
Department of Computer Engineering
Hamedan University of Technology
Hamedan , Iran
jalal.abdollahi66@gmail.com

*Abstract*—Nowadays, using cryptographic systems play an effective role in security and safety technologies. One of the most applied kind of cryptography is Symmetric Cryptography and its applications. New aspects of symmetric Cryptography methodologies and applications has been presented by this paper. Security-based networks and some complex technologies such as RFID and parallel security settings has been introduced by using Symmetric Cryptography is the main base of discussion in this paper. Designing an unique protocol for Symmetric Cryptography in security networks elements is our focus.
Reviewing benefits of using these methodologies has been presented and discussed in this paper.

## I. INTRODUCTION

It is widely recognized that security issues play a crucial role in the majority of computer and communication systems. A central tool for achieving software protection is Cryptography. Cryptographic algorithms are most efficiently implemented in custom hardware than in software running on general purpose processors. Hardware implementations are of extreme importance in case of high performance, security against system intruders and busy systems, where a cryptographic task consumes too much time. Traditional ASIC solutions have the well-known drawback of reduced flexibility compared to software solutions. Since modern security protocols are increasingly becoming algorithm independent, a high degree of flexibility with respect to the cryptographic algorithms is desirable. The security degrees of all the techniques are based on the hardness of mathematical problems. Among them, Elliptic curve cryptography shows a promise to be an alternative of RSA. A promising solution which combines high flexibility with the speed and physical security of traditional hardware is the implementation of cryptographic algorithms on reconfigurable devices such as FPGAs. FPGAs are hardware devices whose function is not fixed and which can be programmed in-system [1].

It is essential for providing some requirements in Hardware and Software systems security. By using Symmetric Cryptography and its aspects, we can design some complex security systems with encryption algorithms.
When we use a complex system like RFID tags, its security must be covered all of its elements.

## II. SYMMETRIC CRYPTOGRAPHY

Symmetric key systems require both the sender and the recipient to have the same key. This key is used by the sender to encrypt the data, and again by the recipient to decrypt the data. Key exchange is clearly a problem. How do you securely send a key that will enable you to send other data securely? If a private key is intercepted or stolen, the adversary can act as either party and view all data and communications. You can think of the symmetric crypto system as akin to the Chubb type of door locks. You must be in possession of a key to both open and lock the door.
Symmetric cryptography uses a single private key to both encrypt and decrypt data. Any party that has the key can use it to encrypt and decrypt data. They are also referred to as block ciphers.
Symmetric cryptography algorithms are typically fast and are suitable for processing large streams of data.
The disadvantage of symmetric cryptography is that it presumes two parties have agreed on a key and been able to exchange that key in a secure manner prior to communication. This is a significant challenge. Symmetric algorithms are usually mixed with public key algorithms to obtain a blend of security and speed.

## III. SYMMETRIC CRYPTOGRAPHY AND ENCRYPTION MODULES ALGORITHMS IN RFID-BASED SYSTEMS

Radio Frequency Identification (RFID) is an emerging technology. The main idea behind it is to attach a so called RFID tag to every object in a particular environment and give a digital identity to all these objects. An RFID tag is a small microchip, with an antenna, holding a unique ID and other information which can be sent over radio frequency. The information can be automatically read and registered by RFID readers. The data received by the RFID reader can be subsequently processed by a back-end database. Figure 1 gives a graphical overview of an RFID system.

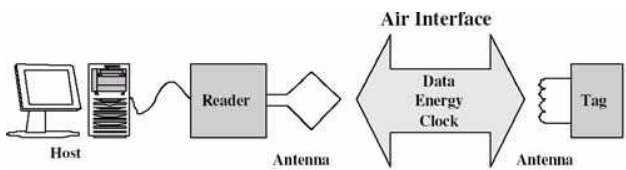

Figure 1: Overview of an RFID system

The range of possible applications varies with the capability of the tag and is separated by different classes. Class 0 and Class 1 RFID tags are used as barcode replacement and are read-only or can be programmed only once in the field, respectively. Inventory maintenance which is used in the supply chain management can be automated using such tags. They are cheap (approximately 5 Cents) and can be used on item-level on nearly every product. This paper is focusing on more advanced tags (Class 2) which also have a rewritable memory and additional hardware resources but do not have an active power supply on the tag. The energy for operation is pulled from the electromagnetic field provided by the reader. In addition, the reader also provides the digital clock frequency for operation. Certain modulation methods are used for communication from the reader to the tag and vice versa. Such tags cost about 50 Cents and the available silicon area is about 10,000 gates.

value products like pharmaceutical and branded goods can be protected against security vulnerabilities. In this paper, we demonstrate how the project ART (Authentication for long-range RFID systems) proposes to improve current RFID systems by providing secure authentication. The project is performed by four independent partners, two from industry and two academic partners. A major goal of the project is to enhance the functionality of current RFID tags with passive power supply. The basic functionality of RFID systems is to provide identification of individual objects by the replies the attached RFID tag sends to a request performed by a reader. The reader uses an attached database to link the received ID number to a specific object described in the database. The major drawback of those systems is that the communication scheme does not provide a method to prove the claimed identity. Since a typical tag answers its ID to any reader (without a possibility to check whether a reader is authorized to receive the information), and the replied ID is always the same, an attacker can easily forge the system by reading out the data of a tag and duplicating it to bogus tags. Closed RFID systems with common access of all readers to a central database, can check for illegal duplicates (bogus tags) within the database but this is not practical for many applications. Furthermore, it is impossible to distinguish the original tag from its illegal duplicates. Strong authentication mechanisms can solve uprising security problems in RFID systems and therefore give protected tags an added value. The three main security threats in RFID systems are forgery of tags, unwanted tracking of customers, and the unauthorized access to the tag's memory. In this paper, we propose authentication protocols for RFID systems based on the ISO/IEC 9798-2 standard [2]. These protocols allow protecting high-value goods against adversary attackers. Additionally, we show that these protocols are feasible for nowadays restriction concerning data rates and compliance to existing standards as well as the requirements concerning chip area and power consumption. With authentication we mean a method to provide a proof for a claimed identity. This proof is based on a secret stored within the authenticating part of the system. As long as the secret information stays secret and the used protocol does not leak sensitive information, an attacker cannot forge a tag. A communication system providing authentication can reject access (to information, entry, etc.) to non authorized parties. To keep the authentication secure, it is necessary that an attacker does not gain information about the secret by listening passively to successful authentications. To fulfill this requirement for strong authentication, it is necessary to use cryptographically strong computations.

*A) Symmetric Authentication*

Authentication is the mechanism that one entity proves its identity to another entity. Strong authentication protocols, such as challenge-response protocols (standardized in ISO/IEC 9798) are widely used in practice today. In challenge-response protocols, one or several messages are exchanged between the party who wants to prove its identity (the claimant) and the party who wants to verify the identity (the verifier). This is called the protocol. In a typical scenario, the verifier challenges the claimant with an unpredictable value that is used no more than once (the nonce). The claimant is required to return a response that is depending on the nonce and on the stored secret. Using strong authentication for RFID systems leads to significant security enhancements. If readers are required to authenticate themselves to tags, attacks such as unwanted tracking and unauthorized memory access are rendered infeasible. If tags are required to authenticate themselves against readers forgery of tags is prevented. It is advantageous to use standardized protocols and algorithms because they have been rigorously cryptanalyzed and are widely used. systems based on standardized protocols and algorithms are more likely to be secure and interoperable with other well established infra-structures. Standardized challenge-response protocols are defined upon symmetric-key and asymmetric-key cryptographic primitives. Using symmetric-key cryptography has the disadvantage that there is one secret key shared through all parties. If one key is compromised for any reason the whole systems gets insecure. However, strong asymmetric-key cryptography requires extremely costly arithmetic operations and is therefore out of question for RFID systems today. Strong symmetric-key cryptographic primitives include encryption primitives such as AES [3] which allow compact implementations [1]. In the following, a few authentication protocols based on challenge-response methods are explained.

*1. Protocol Extension.* The most important command is the anti-collision sequence which is a command every tag has to implement. Thereby, the reader sends an initial inventory command. All tags in the environment make a response which is the tag's unique ID. If only one tag answers to the request the ID can be retrieved by the reader and all subsequent commands can be addressed using the ID which addresses one single tag. If two or more tags make an answer to a request a collision occurs. This can be detected at the reader. The reader then uses a modified inventory request where it adds a part of the tag's ID to the request. Only tags which have this part of the ID are allowed to answer. Once the ID of one tag is identified, the reader sends a "stay quiet" command to the tag with the identified ID. This method is used as long as there are no more collisions and all tags within the environment are identified. Adding an authentication command to the ISO 18000 standard works by using a custom command which can be defined. The challenge-response protocol fits ideally to the overall request-response protocol. When authenticating a tag, the reader sends a challenge within the request and the tag answers according to the presented authentication protocol.

*2. Interleaved Authentication Protocol.* The authentication protocol mentioned above only works when the result of the cryptographic primitive is available within the time defined for the tag's response. As this time is very short a modification of this authentication scheme was proposed where the calculation time for the algorithm is of minor importance. For this purpose, authentication is split into two parts. The first part is the Authentication Request (AR), which tells the tag to encrypt the challenge and does not expect any response. The second part is the Response Request (RR), which collects the authentication response, when the result is available. For one tag, the timing overhead is large, but with more than one tag, the reader can use the idle time (during the tag is busy calculating) to send authentication requests (or other requests) to other tags [4] .

*3.Cryptographic Hardware Module.* Computation of the cryptographic algorithm AES (Advanced Encryption Standard) is computationally very complex compared to other tasks of tags. The implementation of the AES that fulfils the requirements concerning low power consumption and low die size is far away from being trivial. The current consumption of additional hardware components on an RFID tag must not exceed 10μA to avoid reduction of the operating range [5] .

By using these algorithms, we can design a complex encryption systems that can be influenced on software and hardware modules in RFID-based systems. Our main strategy in defining a security base for complex systems is with following steps:

- Software-based Security(SbS)
- Hardware-based Security(HbS)
- Complexity-based Security(CbS)

This methodology is our main focus on "Parallel Cryptography" which has been designed in three separate steps and finally added to an unique cycle that its planning is in Symmetric Cryptographic algorithms as showed in figure 2.

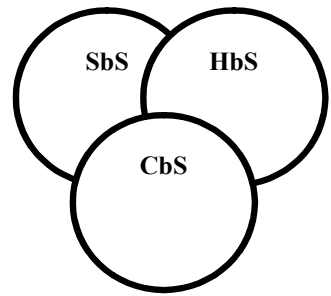

Figure 2 : Complex Symmetric Cryptography diagram

IV. BENEFITS OF PARALLEL CRYPTOGRAPHY USING

When a complex system uses Parallel Security System (PSS), probability of mistakes in description has been decreased, because of designing a complex system with three main security base that can support security of software, hardware and a combination security by Symmetric Cryptography and parallel security. Figure 3 shows that how a Parallel Security System can cover all of cryptographic security base in a complex design.

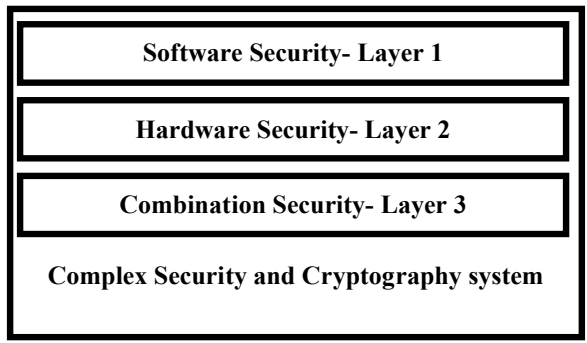

Figure 3. Parallel Security System in a complex design

## V. CONCLUSION

At this paper, we show that the most application Symmetric-based security systems in complex designing is Parallel Security System (PSS). When we use a Parallel Security System, we can support all of software and hardware layers in a complex design and by using this algorithm coverage of all subsystems can be provided. Our main strategy discussion in a sample complex system such as RFID tags can make a secure and safe space for using these kinds of systems. Three steps PSS in complex designing can make a cover layer between processing all of procedures .